\documentstyle{article}

\begin{document}

\title{Hydrodynamic Formulation \\ of the Hubbard Model}
\author{Girish S, Setlur}
\maketitle

\begin{abstract}
 In this article, we show how to recast the Hubbard model in one dimension
 in a hydrodynamic language and use the path integral approach to compute the 
 one-particle Green function.
  We compare with the Bethe ansatz results of
 Schulz and find exact agreement
 with the formulas for spin and charge velocities and anomalous exponent
 in weak coupling regime. 
 These methods may be naturally generalized to more than one dimension by
 simply promoting wavenumbers to wavevectors.
\end{abstract}

\section{Introduction}

 The Hubbard model is one of the most extensively studied models in Condensed 
 Matter Physics. It it the simplest example of an interacting Fermi system on
 a lattice. In one dimension the ground state and excited states may be
 written down explicitly using Bethe ansatz\cite{Lieb}. The collection
 of reprints by Korepin and Essler\cite{Korepin} is particularly useful. 
 The pioneering work of Schulz\cite{Schulz}  in computing the spin and charge
 velocities and the anomalous exponent for all values of the onsite
 repulsion $ U $ is quite significant. There have been other attempts notably
 by Weng et.al.\cite{Weng}
 in recasting the Hubbard model in a path integral language.
 Unfortunately all these methods are restricted to one dimension.
 One notable exception is the work by Liu \cite{Liu} that uses
 eigenfunctional theory. Our approach is quite different from all these
 and uses the hydrodnamic formulation recently developed by the author.
 It it simple and powerful as we shall see since it is naturally generalisable
 to more than one dimension and yields the important correlation functions
 quite easily. Our approach is able to give the right functional dependence of
 the various quantities such as spin and charge velocities and exponents 
 (as a function of $ U $) for small  $ U $. The formalism is naturally
 generalisable to more than one dimension by simply promoting wave numbers to
 wavevectors. Unfortunately we are unable to probe the large $ U $ case 
 which is relevant to high $ T_{c} $ since 
 the formalism does not reproduce the right results in one dimension.

\section{The Hydrodynamic Formulation}

 The program of quantizing hydrodynamics has a long and distinguished history. 
 Landau \cite{Landau} and his students
 were among the first to attempt this. Later
 on Sunakawa et.al. \cite{Sunakawa} and others - notably Rajagopal and Grest
 \cite{Raja} took this program further. Dashen, Sharp, Menikoff and
 Goldin\cite{Sharp} in the seventies introduced many of these ideas.
  Recently, Jackiw and collaborators\cite{Jackiw}
  have revived interest in this approach in the context of relativistic quarks.
 In our earlier work, we introduced the density phase
 variable ansatz for fermions\cite{Setlur1}.
 We also note that Rajagopal and Grest\cite{Raja} had already in the
 seventies pointed out the need
 for having a nonzero-phase functional found in the density phase
 variable ansatz. In our earlier work\cite{Setlur1} we
 made a first pass at computing the phase functional.
 This attempt yielded an answer that in retrospect is quite wrong.
 Upon closer examination 
 the $ U_{0}( {\bf{q}} ) $ of our earlier work\cite{Setlur1} is imaginary
 when it was postulated to be real(for small $ {\bf{q}} $). 
 So far the author has avoided this issue by taking refuge under the
 the sea-boson approach that enables us to derive the momentum distribution,
 anomalous exponents, quasiparticle residue and so on without yielding the
 full dynamical propagator which is of interest
 only because it contains information
 about quantities just mentioned. If one is able to compute them without
 having to compute the full propagator so much the better.
 However, there are physical problems
 in which the full propagator is important.
 The X-ray edge problem\cite{Mahan} is one such.
 In fact we tried using the DPVA to
 compute the X-ray edge spectra in a
 preprint\cite{preprint} and found that we obtain the
 right answers in one dimension but the answers in higher dimensions were
 inconsistent with Mahan's exact results\cite{Mahan}. 

 In an earlier preprint, after much reflection,
 we chose to dismiss the approach that only uses the hydrodynamic
 variables namely the density and its conjugate as `myopic' 
 (mypoic bosonization). This is beacause a hamiltonian formulation in terms
 of the hydrodynamic variables is unable to distinguish between fermions
 and bosons. We have to further decompose these variables in terms of linear
 combination of oscillators in order to distinguish between the two statistics.
 However, the sea-boson approach is not without its share of problems. For one
 it does not generalise to finite temperatures easily. Also the full dynamical
 propagator is not reducible to quadratures due to a technical difficulty.
 Both these problems may be resolved in an approach that incorporates
 only the hydrodynamical variables.  
 We show in this preprint, that the path integral approach is 
 an avenue to distinguish between the statistics when using only the 
 hydrodynamical variables.  
  This approach that only uses hydrodynamic variables
 was not expected to work out for fermions, since one has to take into account
 the extended nature of the Fermi surface. However, this idea seems
 too important to pass up. In particular, the natural manner in which 
 gauge theory may be studied in this approach\cite{Bose} makes this effort 
 for fermions worthwhile and urgent.

 In this section, we introduce the hydrodynamic formulation that has been
 developed by the author. It involves writing the field variable in trems
 of observables such as currents and densities. In the long-wavelength limit,
 it can be shown that current algebra (that is, the mutual commutation rules
 between currents and densities) is obeyed only if the current operator
 is expressible as shown below.
\begin{equation}
{\bf{J}}({\bf{x}},t) = - \rho({\bf{x}},t) \mbox{    }\nabla \Pi({\bf{x}},t)
\end{equation}
 Here $ \Pi $ is the potential for the velocity and is conjuugate
 to the density $ \rho $, Thus in the hydrodynamic limit the velocity operator
 is irrotational. The hydrodynamic part of the field variable
 may be written in a polar form.
\begin{equation}
\psi_{slow}({\bf{x}},t) = e^{ i \Lambda([\rho];{\bf{x}},t) }
 e^{ -i \Pi({\bf{x}},t)  } \sqrt{ \rho({\bf{x}},t) } 
\end{equation}
 For fermions, we expect
 $ \psi_{slow}({\bf{x}},t) $ to be a Grassmann variable. But we shall take the
 point of view that the fermionic nature of the field is captured at the level
 of the propagator by the introduction of the phase functional
 $ \Lambda([\rho];{\bf{x}},t) $. The fermionic KMS boundary conditions
 is obeyed since a position independent global Klein factor is able to capture
 this\cite{Setlur2}. Thus the program involves expanding
 $ \Lambda([\rho];{\bf{x}},t) $ in powers of the density fluctuations and
 making contact with the free theory and fixing the coefficients. 
 The rest of the discussion is similar to our earlier work 
 on bosons\cite{Bose}, therefore we shall not dwell on those details.
 Suffice it to say that if we expand the action in powers of the desnity
 fluctuations and retain only the harmonic parts, in order to
 recover the right desnity-density correlation functions, we have to set,
\begin{equation}
\lambda([\rho];{\bf{q}}n) = C({\bf{q}}n) \mbox{        }
\rho_{ -{\bf{q}}, -n } 
\end{equation}
and $ \Lambda([\rho];{\bf{x}},t) = \sum_{ {\bf{q}},n } e^{ i{\bf{q}}.{\bf{x}}
} e^{ z_{n} t } \lambda([\rho];{\bf{q}}, n) $. The coefficient may be computed
as follows.
\begin{equation}
 \beta z_{n} \mbox{      } C({\bf{q}}n) 
 = \frac{1}{ 2 \left< \rho_{ {\bf{q}}, n } \rho_{ -{\bf{q}}, -n } \right>_{0} }
- \frac{ \beta z^2_{n} }{ 4 N^{0} {\bf{q}}^2 } 
-  \frac{ \beta {\bf{q}}^2 }{ 4 N^{0} } 
\end{equation}
 and $ \left< \rho_{ {\bf{q}}, n } \rho_{ -{\bf{q}}, -n } \right>_{0} $ is
 the density-density correlation function of the free theory obtained from
 elementary considerations.
 In one dimension, for spinless fermions we may write,
\begin{equation}
C(q,n) = \frac{ v^2_{F} }{ 4 N^{0} z_{n} }
\end{equation}
where $ v_{F} $ is the Fermi velocity and $ z_{n} = 2 \pi n/\beta $ is the
 bosonic Matsubara frequency.

\section{Hubbard Model}

 The Hubbard model in one dimension\cite{Mahan}
 may be written down as follows(here $ G = 2\pi/a $).
\begin{equation}
H = -2t \sum_{k \sigma} cos(ka) \mbox{          }
 c^{\dagger}_{k \sigma } c_{ k \sigma }
+ \frac{ U }{ N_{a} } \sum_{q}\rho_{q \uparrow } \rho_{ -q  \downarrow }
+ \frac{ U }{ N_{a} }\sum_{q}\rho_{q \uparrow } \rho_{ -q + G \downarrow }
+ \frac{ U }{ N_{a} } \sum_{q}\rho_{q \uparrow } \rho_{ -q - G \downarrow }
\end{equation}
 First we replace the cosine dispersion by a parabolic one. 
 $ \epsilon_{k} = -2t \mbox{       }cos(ka) = c_{0} + c_{1} k^2 $.
 From this we find $ c_{0} = -2t $ and we require that the slope of
 the dispersion at $ k = \pm k_{F} $ be identical to the cosine dispersion. 
 This means $ c_{1} = (ta/k_{F}) sin(k_{F}a) = 1/(2m)$. 

 Thus we may write a quadratic action
 for the one band Hubbard model in one dimension, including umklapp processes
 in the hydrodynamic language.
\[
S_{Hubb} = \sum_{ q\sigma, n } (-i \beta z_{n}) \rho_{ q\sigma, n } 
X_{ q\sigma, n } 
 + \frac{ i \beta N^{0} }{2} \sum_{ q\sigma, n } \frac{ q^2 }{2m} \mbox{     } 
 X_{ q\sigma, n }  X_{ -q\sigma, -n } 
\]
\[
 + i \beta  \sum_{ \sigma q \neq 0 n }
  \mbox{        }z_{n} C(qn) \mbox{    }
\rho_{ q\sigma, n }
\rho_{ -q\sigma, -n } 
+ \frac{ i \beta U }{ N^{0} }  \sum_{ q \neq 0 n }
\rho_{ q \uparrow, n } \rho_{ -q \downarrow,-n}
\]
\begin{equation}
+ \frac{ i \beta U }{ N^{0} }  \sum_{ q \neq 0 n }
\rho_{ q \uparrow, n } \rho_{ -q+G \downarrow,-n}
+ \frac{ i \beta U }{ N^{0} }  \sum_{ q \neq 0 n }
\rho_{ q \uparrow, n } \rho_{ -q-G \downarrow,-n}
\end{equation}
 where $ G = 2\pi/a $. The quadratic action implicitly ignores
 three body density correlation functions.
 It is not clear why this is valid 
 except that it renders the path integrals tractable.
 However we may expect to find nontrivial results already at the
 harmonic level with umklapp processes for strong coupling.
 Since we are considering umklapp process which involves large momentum
 transfer, we have to make sure that the $ C(qn) $ is evaluated in general
 for both small and large $ q $. In our earlier work we showed,
\begin{equation}
 \beta z_{n} \mbox{      } C(q\sigma,n) 
 = \frac{1}{ 2 \left< \rho_{ q\sigma, n } \rho_{ -q\sigma, -n } \right>_{0} }
- \frac{ (2m)\beta z^2_{n} }{ 2 N^{0} q^2 } 
-  \frac{ \beta q^2 }{ 2 N^{0}(2m) } 
\end{equation}
 where $ N^{0} $ is the total number of electrons including both spins
 and $ \left< \rho_{ q\sigma, n } \rho_{ -q\sigma, -n } \right>_{0} $ is
 the density-density correlation function of the free theory evaluated using
 elementary considerations.
 The field variable is now given by,
\begin{equation}
\psi_{slow}(x\sigma,t) 
= e^{ -i \sum_{q,n} e^{ iqx } e^{z_{n} t} X_{q\sigma,n} }
 \mbox{     } e^{ \eta \mbox{   }
i \sum_{q,n} e^{ iqx } e^{z_{n} t} C(q\sigma,n) 
\rho_{-q\sigma,-n} }
\end{equation}
Here $ \psi_{slow} $ is the hydrodynamic (slow) part of the field.
Also $ \eta $ is a `fudge factor' needed to make sure that we 
recover the right exponents. 
It is a numerical factor like $ 1/2 $ or $ 2 $. 
In the first instance, we may neglect umklapp processes.
 This corresponds to weak coupling and small $ q $.
\[
\left< T \psi(x,\uparrow,t)  \psi^{\dagger}(x^{'},\uparrow,t^{'})  \right> = \int D[\rho_{ \downarrow } ]
e^{ -\sum_{ q, n} {\tilde{\lambda}}_{q,n} \rho_{q,\downarrow,n}
\rho_{-q,\downarrow,-n} }
e^{ -\sum_{q,n}  b^{0}_{q,n} 
  \frac{ \beta U }{ 2 \lambda_{q,n} N^{0} } \rho_{ q \downarrow, n }  }
\]
\[
b^{0}_{q,n} = \left[ \frac{1}{ \beta N^{0} } \frac{2m}{q^2} 
(\beta z_{n}) (-i) 
 + \eta \mbox{   }i \mbox{   }C(-q,-n) \right]
\left( e^{-iqx} e^{ -z_{n}t } -  e^{-iqx^{'}} e^{ -z_{n}t^{'} } \right)
\]
\[
{\tilde{\lambda}}_{q,n} = \lambda_{q,n}
 - \left( \frac{ \beta U }{ N^{0} } \right)^2/(4 \lambda_{q,n})
\]
\[
\lambda_{q,n} = \left( \frac{1}{ 2 \beta N^{0} } \right)
\left( \frac{2m}{ q^2 } \right)
(\beta z_{n})^2  + (\beta z_{n}) \mbox{   }C(q,n)
\]
\[
C(q,n) = \frac{ 4 \epsilon_{F} }{ N^{0} z_{n} }
\]

\begin{equation}
\left< T \psi(x,\uparrow,t) \psi^{\dagger}(x^{'},\uparrow,t^{'}) \right> = 
e^{ \sum_{q, n} \frac{ b^{0}_{q,n} b^{0}_{-q,-n} }
{ 4  \lambda_{q,n}
 - \left( \frac{ \beta U }{ N^{0} } \right)^2/\lambda_{q,n} } 
\left( \frac{ \beta U }{ 2 \lambda_{q,n} N^{0} } \right)^2 }
\end{equation}
If we set $ \eta = 2 $ we find,
\[
b^{0}_{q,n} = -\frac{ 2i }{ \beta z_{n} }
\mbox{   }\lambda_{q,n} \mbox{    }
\left( e^{-iqx} e^{ -z_{n}t } -  e^{-iqx^{'}} e^{ -z_{n}t^{'} } \right)
\]
\begin{equation}
\left< T \psi(x,\uparrow,t) \psi^{\dagger}(x^{'},\uparrow,t^{'}) \right> = 
e^{ \sum_{q, n} \frac{1}{ z_{n}^2 }
\mbox{   }\frac{ 4 \lambda_{q,n} }{ 4 \lambda^2_{q,n}
 - \left( \frac{ \beta U }{ N^{0} } \right)^2 }\mbox{    }
\left( 2 - e^{-iq(x-x^{'})} 
e^{ -z_{n}(t-t^{'}) } -  e^{iq(x-x^{'})} e^{ z_{n}(t-t^{'}) } \right) 
\left( \frac{ U }{ 2 N^{0} } \right)^2 }
\end{equation}
For weak coupling we have,
\[
\left< \psi^{\dagger}(x^{'},\uparrow,t^{'}) \psi(x,\uparrow,t) \right> \approx
e^{ - \frac{ \pi U^2 }{ 4 k_{F} }
 \sum_{j=1,2} \frac{ 1 }{ (2 \pi)  4 m v_{F}^3 } \int^{ \infty }_{ 0 } 
 \frac{ dq }{ |q| } \mbox{    }
\left( 2 - e^{-iq [ (x-x^{'}) - v_{F,j} (t-t^{'}) ] } 
 -  e^{iq [ (x-x^{'}) + v_{F,j} q (t-t^{'}) ] } \right)  }
\]
\begin{equation}
 = \prod_{ j = 1, 2 }
 \left( \frac{1}{ [ (x-x^{'}) -   v_{F,j} (t-t^{'}) ]^{ \alpha } } \right)
 \left( \frac{1}{ [ (x-x^{'}) +  v_{F,j} (t-t^{'}) ]^{ \alpha } } \right)
\end{equation}
Here, $ v_{F,j} = v_{F} \left( 1 \pm \frac{ U }{ 2 m v^2_{F} } \right)^{\frac{1}{2}} $.
The anomalous exponent related to the momentum distribution is given by,
\begin{equation}
\gamma = 2\alpha = U^2  \frac{ 1 }{ 16 ( k_{F}^2/m)^2 }
\end{equation}
 The Bethe ansatz  result of Schulz shows that
  $ \gamma = U^2/(4 \pi^2 v_{F}^2) $. In units such that
 $ a = 1 $ near half filling we have $ k_{F} = \pi/2 $ and $ m = \pi/4 $.
 Thus in our case we have, $ \gamma = U^2  \frac{ 1 }{ 16 \pi^2 } $. 
 From the result of
 Schulz also we find, $ \gamma = U^2/(16 \pi^2) $.
 Therefore this approach gives the right spin and charge velocities
 (especially close to half-filling), 
 and the right anomalous exponent for small $ U $.

 However, for large $ U $, this approach does not give us the right qualitative
 behaviour as it predicts $ \gamma  \sim \sqrt{ U } $. 
 According to the Bethe ansatz solution the anomalous exponent saturates 
 to a value $ \gamma = 1/8 $. 
 For strong coupling, we are unable to make progress. We have tried using the
 no-double occupany constraint since this is easy to implement using the
 hydrodynamic formulation, but we have been unable to make it work. 
 The umklapp terms are also unfortuantely of little
 use. Thus we shall not advocate the use of this approach for strong coupling.
 In two spatial dimensions we expect to find that the system is a Landau Fermi
 liquid. We shall not carry out this calculation since it is not
 very interesting. The main purpose of this article is to highlight the
 usefullness of the hydrodynamic
 approach and its generality so that in future publications we may use this
 to study disordred systems and the like.

 This work was supported by the Harish Chandra Research Institute.

\end{document}